\documentclass[floatfix,aps,prd,amsmath,nofootinbib,preprintnumbers,twocolumn]{revtex4}

\usepackage{graphicx}
\usepackage{bm}

\def\half{{1\over 2}}

\def\half{{1\over 2}}
\def\({\left(}
\def\){\right)}
\def\[{\left[}
\def\]{\right]}

\def\e{\begin{equation}}
\def\q{\end{equation}}
\def\m{\begin{eqnarray}}
\def\n{\end{eqnarray}}

\begin{document}

\title{Inflation model selection revisited}

\author{Jun Li$^{1,2}$\footnote{lijun@itp.ac.cn} and Qing-Guo Huang$^{1,2}$ \footnote{huangqg@itp.ac.cn}}
\affiliation{$^1$ CAS Key Laboratory of Theoretical Physics,\\ Institute of Theoretical Physics, \\Chinese Academy of Sciences, Beijing 100190, China\\
$^2$ School of Physical Sciences, \\University of Chinese Academy of Sciences,\\ No. 19A Yuquan Road, Beijing 100049, China\\
}

\date{\today}

\begin{abstract}

We update the constraints on the power spectra of primordial curvature perturbation and tensor perturbation including Planck data 2015 (P15) and recently released BICEP2/Keck data (BK15), Baryon Acoustic Oscillation data and the Type Ia supernovae data. We find that the upper limits of tensor-to-scalar ratio are $0.061$, $0.064$ and $0.072$ at $95\%$ confidence level (CL) in the $\Lambda$CDM+$r$, $\Lambda$CDM+$r$+$\alpha_s$ and $\Lambda$CDM+$r$+$\alpha_s$+$\beta_s$ models respectively, where $\alpha_s$ and $\beta_s$ are the running of scalar spectral index and running of running. The inflation model with a concave potential is favored at more than $95\%$ CL. In addition, parametrizing the slow-roll parameter $\epsilon\sim 1/N^p$, where $N$ is the e-folding number before the end of inflation and taken in the range of $[50,60]$ and $[14,75]$ respectively, we conclude that the inflation model with a monomial potential $V(\phi)\sim \phi^n$ is disfavored at more than $95\%$ CL, and both the Starobinsky inflation model and brane inflation model are still consistent with the data.

\end{abstract}

\pacs{???}

\maketitle


\section{Introduction}

Inflation \cite{Starobinsky:1980te,Guth:1980zm,Linde:1981mu,Albrecht:1982wi} is proposed to be the elegant paradigm for the very early universe. Not only can it easily solve all of the puzzles, such as the flatness problem, horizon problem and so on, in the hot big bang model, but also provides primordial density perturbation for seeding the temperature anisotropies in the cosmic microwave background (CMB) and formation of the large-scale structure in our Universe. Furthermore, as the fundamental degree of freedom of gravity, gravitational waves are also excited during inflation and finally leave a fingerprint (B-mode polarization) in CMB. 
Even though BICEP2 collaboration detected the excess of B-mode power over the base lensed-$\Lambda$CDM expectation \cite{Ade:2014xna} in the early of 2014, such a signal is finally explained by the thermal dust, not the primordial gravitational waves  and can be explained by the polarized thermal dust, not the primordial gravitational waves in \cite{Cheng:2014pxa} which was confirmed by a joint analysis of B-mode polarization data of BICEP2/Keck Array and \textit{Planck} (BKP) in \cite{Ade:2015tva}. 

Since inflation happened in the very early universe, we can learn it from the measurement of cosmic structure, in particular from the CMB temperature anisotropies and polarizations including Planck satellite \cite{Ade:2015xua,Ade:2015lrj,Aghanim:2016yuo}, BICEP2 and Keck observations through 2015 reason \cite{Ade:2018gkx}. In order to achieve a better constraint on the power spectra of primordial curvature perturbation and tensor perturbation, we also consider to use Baryon Acoustic Oscillation (BAO) data, including 6dFGS \cite{Beutler:2011hx}, MGS \cite{Ross:2014qpa}, BOSS $\mathrm{DR11}\_{\mathrm{Ly}\alpha}$ \cite{Delubac:2014aqe}, BOSS DR12 with nine anisotropic measurements \cite{Wang:2016wjr} and eBOSS DR14 \cite{Ata:2017dya}, and Type Ia supernovae data (JLA) \cite{Scolnic:2017caz} to constrain the low redshift expansion history of the Universe. 

In this paper we will revisit the previous results in \cite{Huang:2007qz,Huang:2015cke,Li:2018iwg} and adopt the Planck data (P15), BICEP2/Keck data 2015 season (BK15), BAO and JLA datasets to constrain the power spectra of primordial curvature perturbation and tensor perturbation, and then give the latest constraints on the inflation models. Our paper will be organized as follows. In Sec.~II, we parametrize the power spectra of primordial curvature perturbation and tensor perturbation, and compare the constraints on the cosmological parameters in different cosmological models. In Sec.~III, we figure out the latest constraints on the inflation models. The summary and discussion are included in Sec.~IV.

\section{Constrains on the spectral running}

In this section the power spectra of the primordial curvature perturbation and tensor perturbation  are parameterized by 
\m
P_s(k)&=&A_s\(\frac{k}{k_*}\)^{n_s-1+\frac{1}{2}\alpha_s\ln(k/k_*)+\frac{1}{6}\beta_s(\ln(k/k_*))^2+...}, \label{eqs:spectrumscalar}\\
P_t(k)&=&A_t\(\frac{k}{k_*}\)^{n_t+...},\label{eqs:spectrumtensor}
\n
where $k_*$ is the pivot scale which is set as $k_*=0.002$ Mpc$^{-1}$ in this paper, $A_s$ and $A_t$ are the amplitudes of the power spectra of curvature perturbation and tensor perturbation at the pivot scale, $\alpha_s\equiv\mathrm{d} n_s/\mathrm{d}\ln k$ is the running of scalar spectral index, $\beta_s\equiv{\mathrm{d}^2n_s}/{\mathrm{d}\ln k^2}$ is the running of running of scalar spectral index, and $n_t$ is the tensor spectral index. 
In literature, the tensor-to-scalar ratio $r$ is used to quantify the tensor amplitude compared to the scalar amplitude at the pivot scale, namely 
\e
r\equiv\frac{A_t}{A_s}.
\q
For the canonical single-field slow-roll inflation model, $n_t$ is not a free parameter, and is related to $r$ by $n_t=-r/8$ which is also called consistency relation \cite{Liddle:1992wi,Copeland:1993ie}.

In this paper we adopt publicly available codes Cosmomc \cite{Lewis:2002ah} to globally fit the parameters including the other parameters in the standard $\Lambda$CDM model: the baryon density parameter $\Omega_b h^2$, the cold dark matter density $\Omega_c h^2$, the angular size of the horizon at the last scattering surface $\theta_\text{MC}$, the optical depth $\tau$. 
Our results are summarized in Tab.~\ref{table:spectra}. 
\begin{table*}
\newcommand{\tabincell}[2]{\begin{tabular}{@{}#1@{}}#2\end{tabular}}
  \centering
  \begin{tabular}{c | c | c | c}
  \hline
  \hline
  Parameters & \tabincell{c}{$\Lambda$CDM+r}& \tabincell{c}{$\Lambda$CDM+r+$\alpha_s$}  & \tabincell{c}{$\Lambda$CDM+r+$\alpha_s$+$\beta_s$}\\
  \hline
  $\Omega_bh^2$ &  $0.02233\pm0.00013$ & $0.02234\pm0.00014$ & $0.02229\pm0.00014$\\
  $\Omega_ch^2$ &   $0.1182\pm0.0007$ & $0.1182\pm0.0007$ & $0.1183\pm0.0007$\\
  $100\theta_{\mathrm{MC}}$ &  $1.0410\pm0.0003$ & $1.0410\pm0.0003$ & $1.0410\pm0.0003$\\
  $\tau$ &  $0.062\pm0.008$ & $0.062\pm0.008$ & $0.063\pm0.007$\\
  $\ln\(10^{10}A_s\)$  & $3.159\pm0.017$ & $3.149\pm0.034$ &$3.123\pm0.040$\\
  $n_s$ &  $0.9680\pm0.0034$ & $0.9748^{+0.0211}_{-0.0214}$ & $1.0193^{+0.0442}_{-0.0440}$\\
  $r_{0.002}$ ($95\%$ CL) &  $<0.061 $ & $<0.064$  & $<0.072$\\
  $\alpha_s$ &  $...$ & $-0.0022^{+0.0069}_{-0.0068}$ & $-0.0401^{+0.0343}_{-0.0338}$\\
  $\beta_s$ &  $...$ & $...$ & $0.0145^{+0.0127}_{-0.0131}$\\
  \hline
  \hline
  \end{tabular}
  \caption{The $68\%$ limits on the cosmological parameters in the $\Lambda$CDM+$r$ model, the $\Lambda$CDM+$r+\alpha_s$ model and the $\Lambda$CDM+$r+\alpha_s+\beta_s$ model from the data combinations of Planck15+BK15+BAO+JLA. }
  \label{table:spectra}
\end{table*}

\begin{figure}
\centering
\includegraphics[width=8cm]{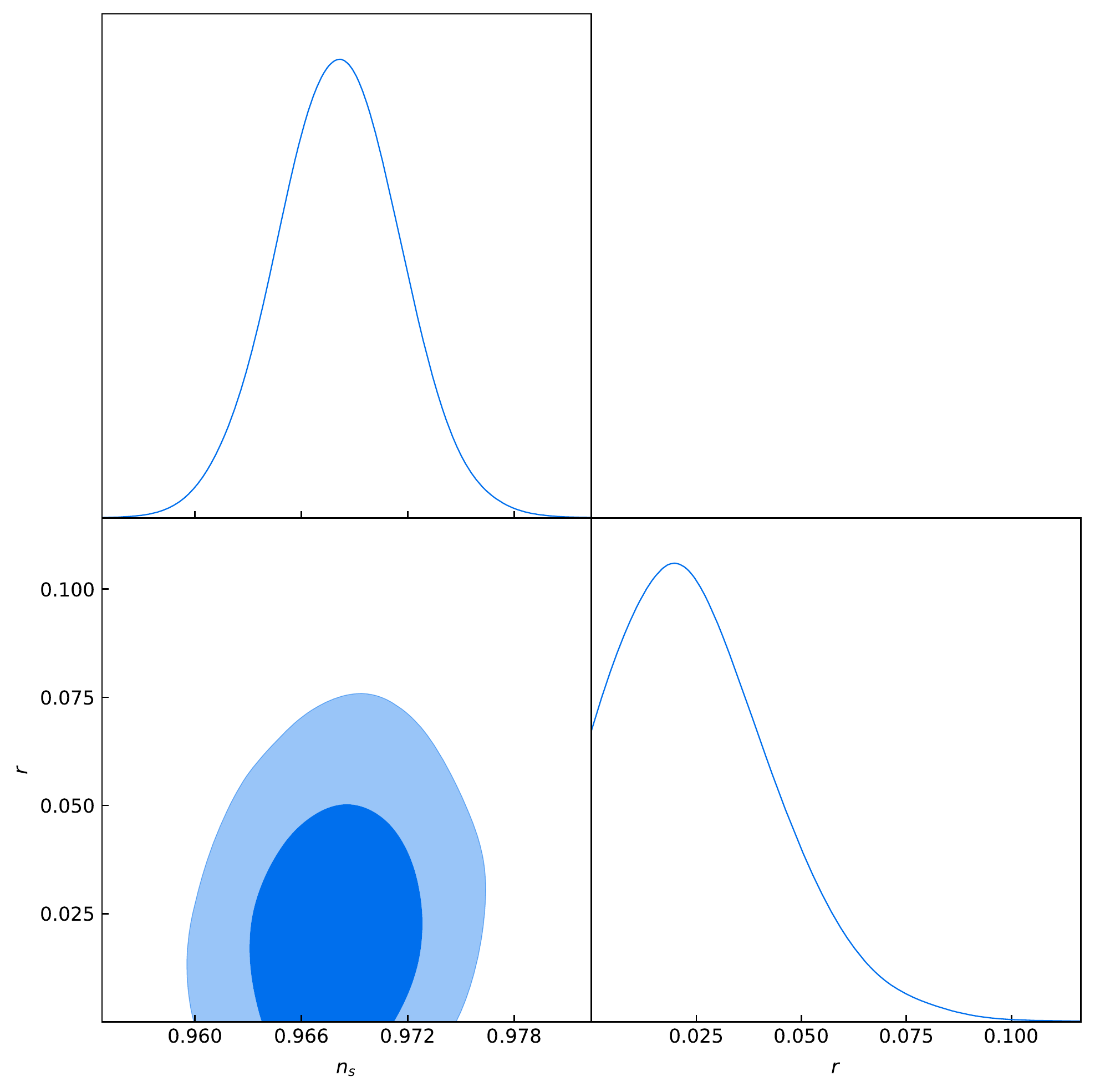}
\caption{The marginalized contour plots and likelihood distributions for parameters $r$ and $n_s$ at $68\%\ \mathrm{CL}$ and $95\%\ \mathrm{CL}$ from Planck15+BK15+BAO+JLA datasets.}
\label{fig:1}
\end{figure}

In the $\Lambda$CDM+$r$ model, the constraints on $r$ and $n_s$ from from Planck15+BK15+BAO+JLA datasets are given by 
\m
r &<& 0.061\quad(95\%\ \mathrm{CL}),\\
n_s &=& 0.9680\pm0.0034\quad(68\%\ \mathrm{CL}). 
\n
The marginalized contour plots about $r$ and $n_s$ is showed in Fig.~\ref{fig:1}. 
It indicates that the power spectrum of curvature perturbation deviates from the exact scale-invariant power spectrum at more than 9$\sigma$ CL.

\begin{figure}
\centering
\includegraphics[width=8.cm]{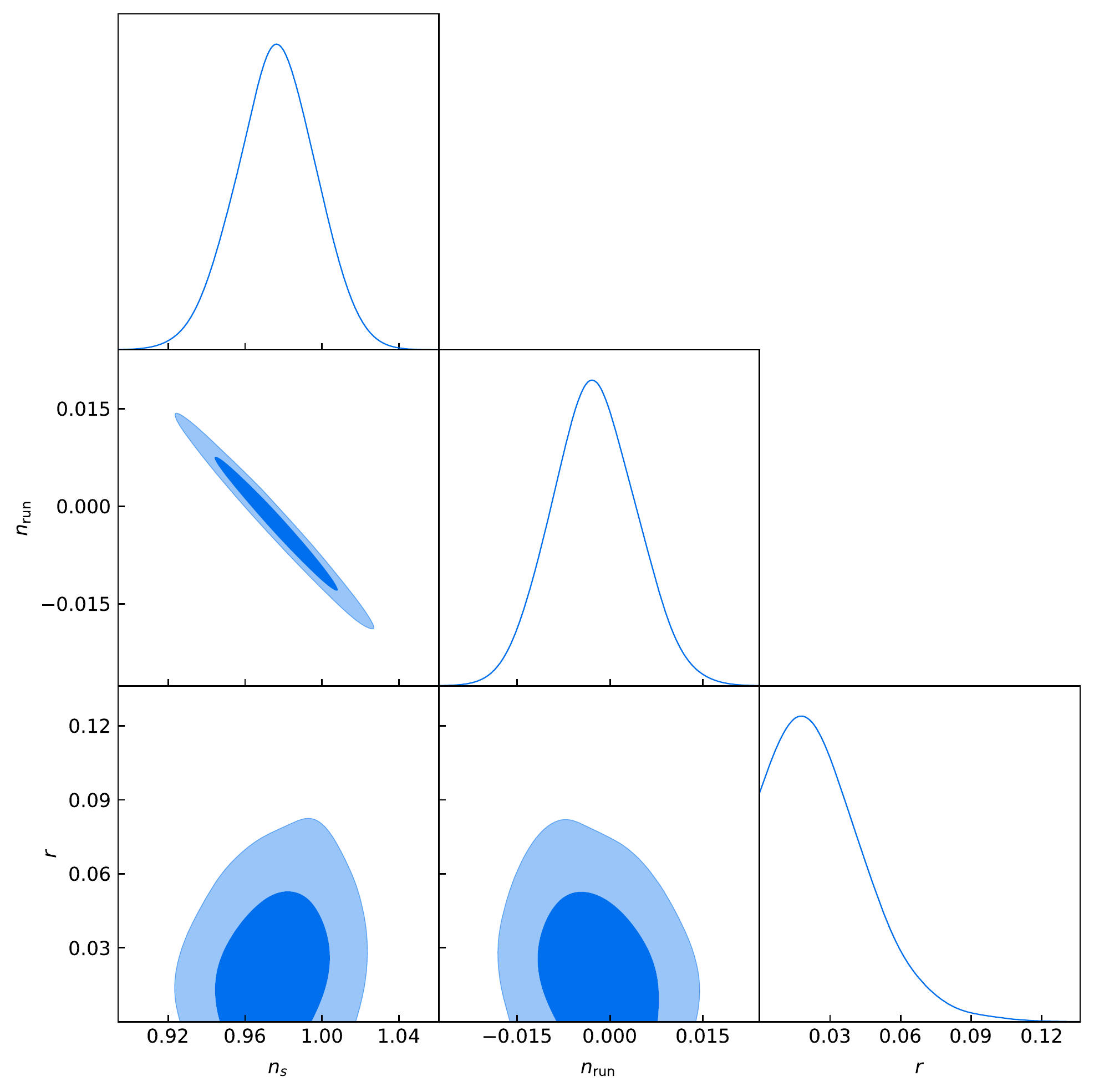}
\caption{The marginalized contour plots and likelihood distributions for parameters $r$, $n_s$ and $\alpha_s$ at $68\%$ CL and $95\%$ CL from Planck15+BK15+BAO+JLA datasets.}
\label{fig:2}
\end{figure}

In the $\Lambda$CDM+$r$+$\alpha_s$ model, the scalar spectral index is running and the constraints on $r$, $n_s$ and $\alpha_s$ from Planck15+BK15+BAO+JLA are
\m
r &<& 0.064 \quad(95\% \ \mathrm{CL}),\\
n_s &=& 0.9748^{+0.0211}_{-0.0214} \quad(68\%\ \mathrm{CL}),\\
\alpha_s &=& -0.0022^{+0.0069}_{-0.0068} \quad(68\%\ \mathrm{CL}). 
\n 
See Fig.~\ref{fig:2} for the the marginalized contour plots and likelihood distributions for parameters $r$, $n_s$ and $\alpha_s$.
A scalar spectral index without running is consistent with the data quite well. 

\begin{figure}
\centering
\includegraphics[width=8.cm]{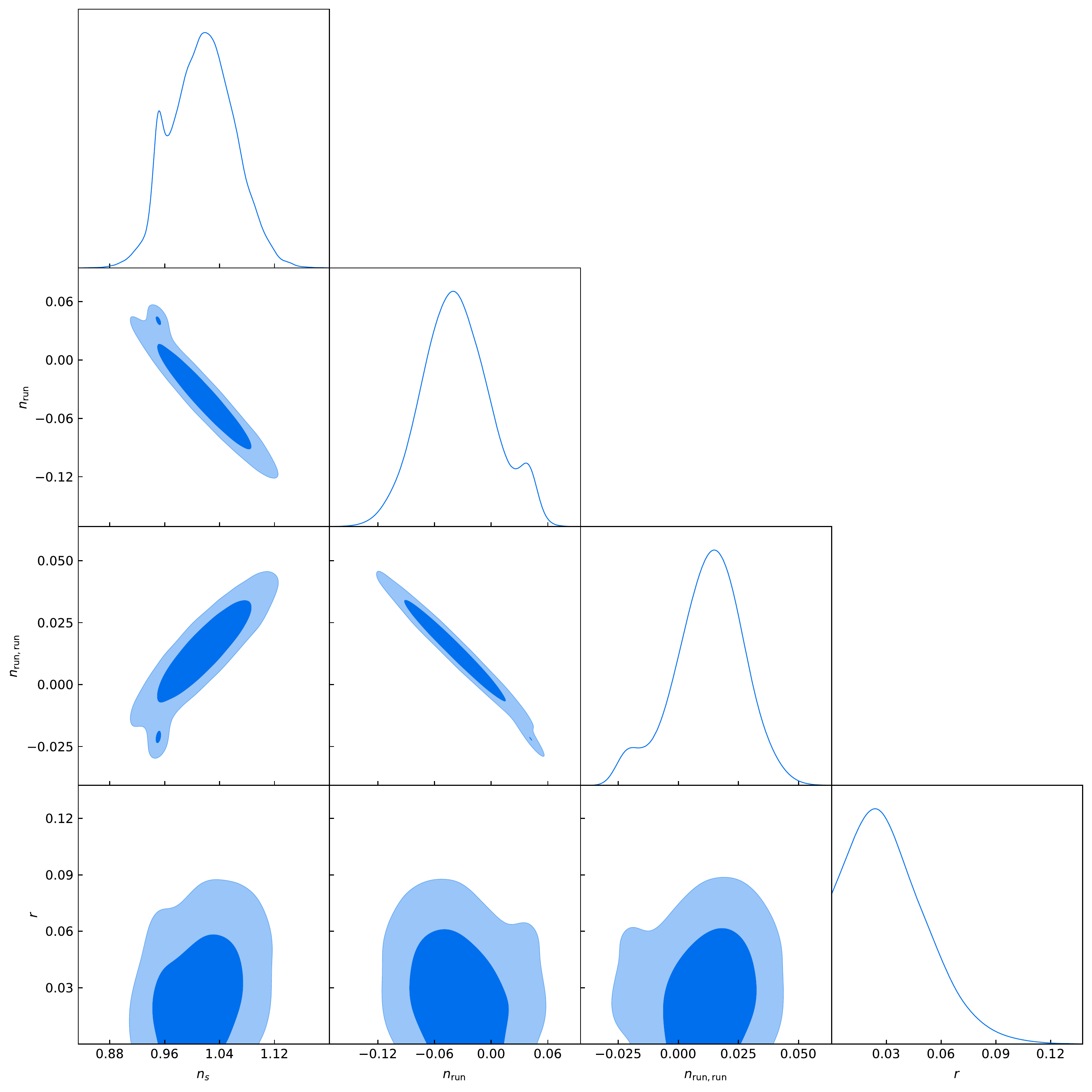}
\caption{The marginalized contour plots and likelihood distributions for parameters $r$, $n_s$, $\alpha_s$ and $\beta_s$ at $68\%\ \mathrm{CL}$ and $95\%\ \mathrm{CL}$ from Planck15+BK15+BAO+JLA datasets.}
\label{fig:3}
\end{figure}

Finally, we extend the former discussion to the model with running of running of scalar spectral index, and the constraints on $r$, $n_s$, $\alpha_s$ and $\beta_s$ from Planck15+BK15+BAO+JLA become 
\m
r &<& 0.072 \quad(95\% \ \mathrm{CL}),\\
n_s &=& 1.0193^{+0.0442}_{-0.0440}\quad(68\%\ \mathrm{CL}),\\
\alpha_s &=&  -0.0401^{+0.0343}_{-0.0338} \quad(68\%\ \mathrm{CL}),\\
\beta_s &=&  0.0145^{+0.0127}_{-0.0131} \quad(68\%\ \mathrm{CL}). 
\n
See the marginalized contour plots for these parameters in Fig.~\ref{fig:3}.
In this model, the constraint on the tensor-to-scalar ratio is slightly relaxed to be $r<0.072$ at $2\sigma$ CL.

\section{Constrains on the inflation models} 

In this section, we will use the observational data to constrain the canonical single-field slow-roll inflation model whose dynamics is govern by 
\m
H^2={1\over 3 M_p^2}\[\half {\dot\phi}^2+V(\phi)\], \\
\ddot \phi+3H\dot\phi+V^\prime (\phi)=0,
\n
where $M_p=1/\sqrt{8\pi G}$ is the reduced Planck energy scale, and the dot and prime denote the derivatives with respective to the cosmic time $t$ and the inflation field $\phi$ respectively. The inflaton field slowly rolls down its potential if  
$\epsilon\ll 1$ and $|\eta|\ll 1$, where the slow-roll parameters $\epsilon$ and $\eta$ are defined by 
\m
\epsilon&=&\frac{M_p^2}{2}\(\frac{V^\prime\(\phi\)}{V\(\phi\)}\)^2,\\
\eta&=&M_p^2\frac{V^{\prime\prime}\(\phi\)}{V\(\phi\)}.
\n
And then the tensor-to-scalar ratio $r$ and the scalar spectral index read 
\m
r&=&16\epsilon,\\
n_s&=&1-6\epsilon+2\eta. 
\n

\subsection{Inflation model selection}

\begin{figure}
\centering
\includegraphics[width=8cm]{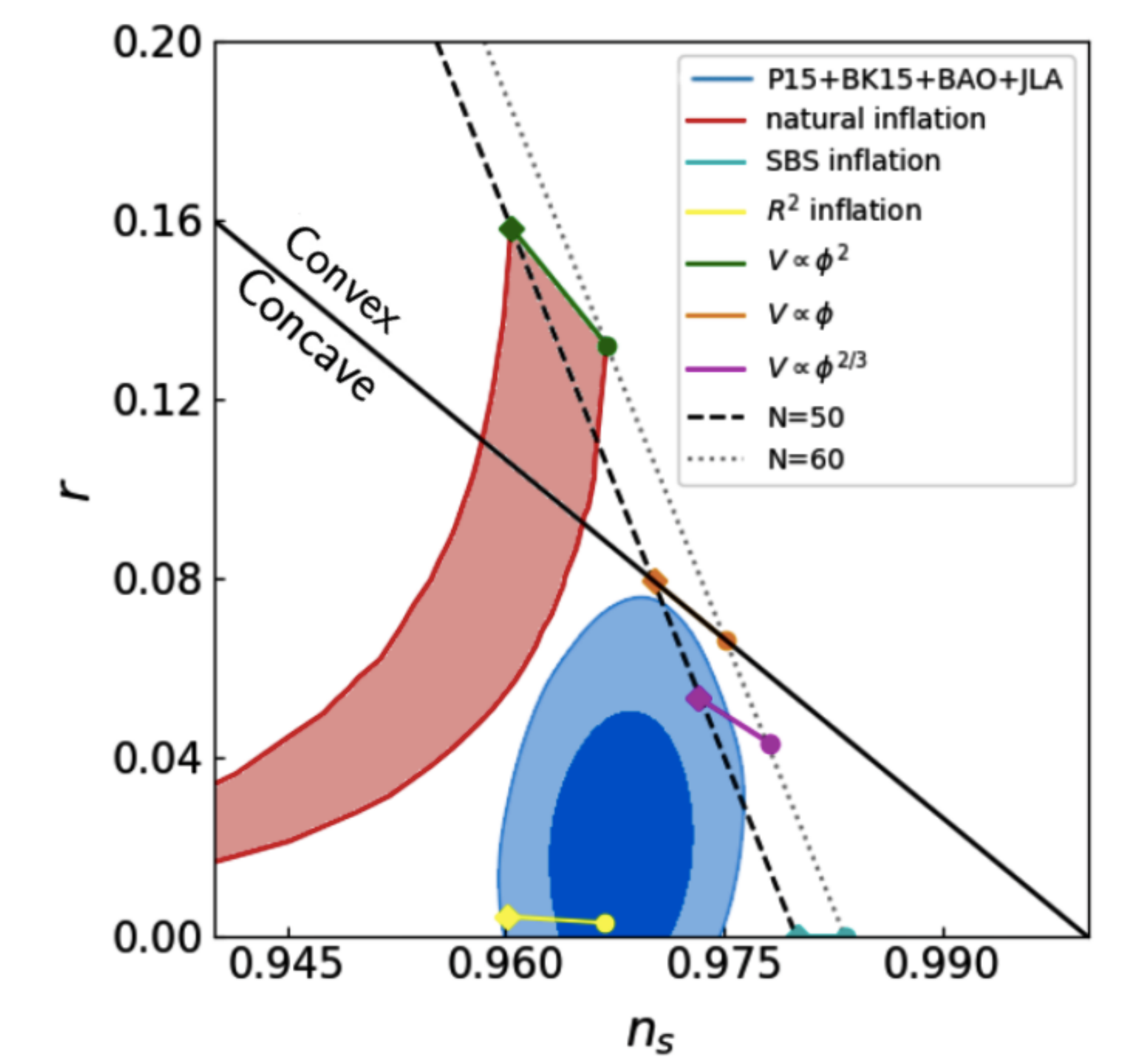}
\caption{The marginalized contour plot for parameters $n_s$ and r at $68\%\ \mathrm{CL}$ and $95\%\ \mathrm{CL}$ from Planck15+BK15+BAO+JLA datasets. The red region represents natural inflation; The cyan line represents SBS inflation; The yellow line represents Starobinsky inflation;  The green line, the orange line and the magenta line represents $\phi^2$, $\phi$ and $\phi^{2/3}$ inflation model.}
\label{fig:4}
\end{figure}

In this subsection we will compare the contour plot of $r-n_s$ constrained by cosmological datasets with the predictions of some simple inflation models. Our main results are showed in Fig.~\ref{fig:4} in which the inflation model with a concave potential is preferred at more than $95\%$ CL.
The detail will be explained in the following part of this subsection. 

The inflation model with a monomial potential $V(\phi)\sim \phi^n$ \cite{Linde:1983gd} is the simplest inflation models. The predictions of this model are given by 
\m
r&=&\frac{14n}{N}, \\
n_s&=&1-\frac{n+2}{2N}.
\n
Here $N$ is the number of e-folds before the end of inflation, and $n$ is not necessarily an integer. For example, axion monodromy in string theory was supposed to realize  $V(\phi)\sim \phi^n$ with $n=2/5$, $2/3$ in \cite{Silverstein:2008sg}, $n=1$ in \cite{McAllister:2008hb}, and the models with higher power in \cite{Marchesano:2014mla,McAllister:2014mpa}. For $N\in [50,60]$, the predictions of inflation model with $V(\phi)\sim \phi^n$ are illustrated in the region between the grey dashed line and the black dashed line in Fig.~\ref{fig:4} which indicates that class of inflation models are marginally disfavored at $95\%$ CL.

The natural inflation model \cite{Freese:1990rb,Adams:1992bn} is govern by the effective one-dimensional potential $V(\phi)=m^2f^2\left(1+\cos(\phi/f)\right)$ where $f$ is the decay constant and $m$ is the mass around local minimum. The tensor-to-scalar ratio and the scalar spectral index for the natural inflation takes the form 
\m
r&=&\frac{8}{(f/M_p)^2}\frac{1+\cos\theta_N}{1-\cos\theta_N}, \\
n_s&=&1-\frac{1}{(f/M_p)^2}\frac{3+\cos\theta_N}{1-\cos\theta_N},
\n
where
\m
\cos\frac{\theta_N}{2}=\exp\left(-\frac{N}{2(f/M_p)^2}\right)\ .
\n
For $50<N<60$, the different decay constant corresponds to different predictions. See the red shaded region in Fig.~\ref{fig:4} which implies that the natural inflation model is strongly disfavored at more than $95\%$ CL.

The spontaneously broken SUSY (SBS) inflation model \cite{SUSY} is proposed to be dominated by the potential   $V(\phi)=V_0\left(1+c\ln\frac{\phi}{Q}\right)$, where $V_0$ is dominant and $c<<1$. The tensor-to-scalar ratio and the scalar spectral index predicted by this inflation model are 
\m
r&\simeq&0, \\
n_s&=&1-\frac{1}{N}.
\n
This model is also disfavored at more than $95\%$ CL.

Starobinsky inflation model \cite{Starobinsky:1980te} is supposed to be driven by a higher Ricci scalar term in the action, namely $S=\frac{M_p^2}{2}\int d^4x\sqrt{-g}\left(R+\frac{R^2}{6M^2}\right)$, where $M$ denotes an energy scale. The tensor-to-scalar ratio and the scalar spectral index in Starobinsky inflation model are 
\m
r&\simeq&\frac{12}{N^2}, \\
n_s&=&1-\frac{2}{N},
\n
in \cite{Mukhanov:1981xt,Starobinsky:1983zz}. Even though this inflation model can fit the data, why the terms with higher powers of Ricci scalar $R$ are all suppressed \cite{Huang:2013hsb} is still an open question.

\subsection{Constraint on typical inflation model}

Because the reheating after the end of inflation is not clear, the exact predictions of inflation models are still unknown due to the uncertainty of the exact number of e-folds before the end of inflation corresponding to the pivot scale $k_*$. In order to solve this problem, similar to \cite{Huang:2007qz,Huang:2015cke}, we can parameterize the slow-roll parameter $\epsilon$ as a function of the e-folding number N before the end of inflation, namely 
\m
\epsilon=\frac{c/2}{\left(N+\Delta N\right)^p},
\n
where c and p are two constant parameters, and
\m
\Delta N=\left(\frac{c}{2}\right)^{1/p}.
\n
And then the tensor-to-scalar ratio and the scalar spectral index take the form 
\m
r&=&\frac{8c}{\left(N+\Delta N\right)^p},\\
n_s&=&1-\frac{c}{\left(N+\Delta N\right)^p}-\frac{p}{N+\Delta N}.
\n
This parameterization can cover many well-known inflation models. For example, $p=1$ and $c=n/2$ for $V(\phi)\sim \phi^n$, $p=2$ and $c=3/2$ for the Starobinsky inflation model, and $p=2(d-1)/d$ and $c\simeq 0$ for the brane inflation model \cite{Dvali:1998pa,Kachru:2003sx} with potential $V(\phi)=V_0(1-(\mu/\phi)^{d-2})$.

\begin{figure}
\centering
\includegraphics[width=8cm]{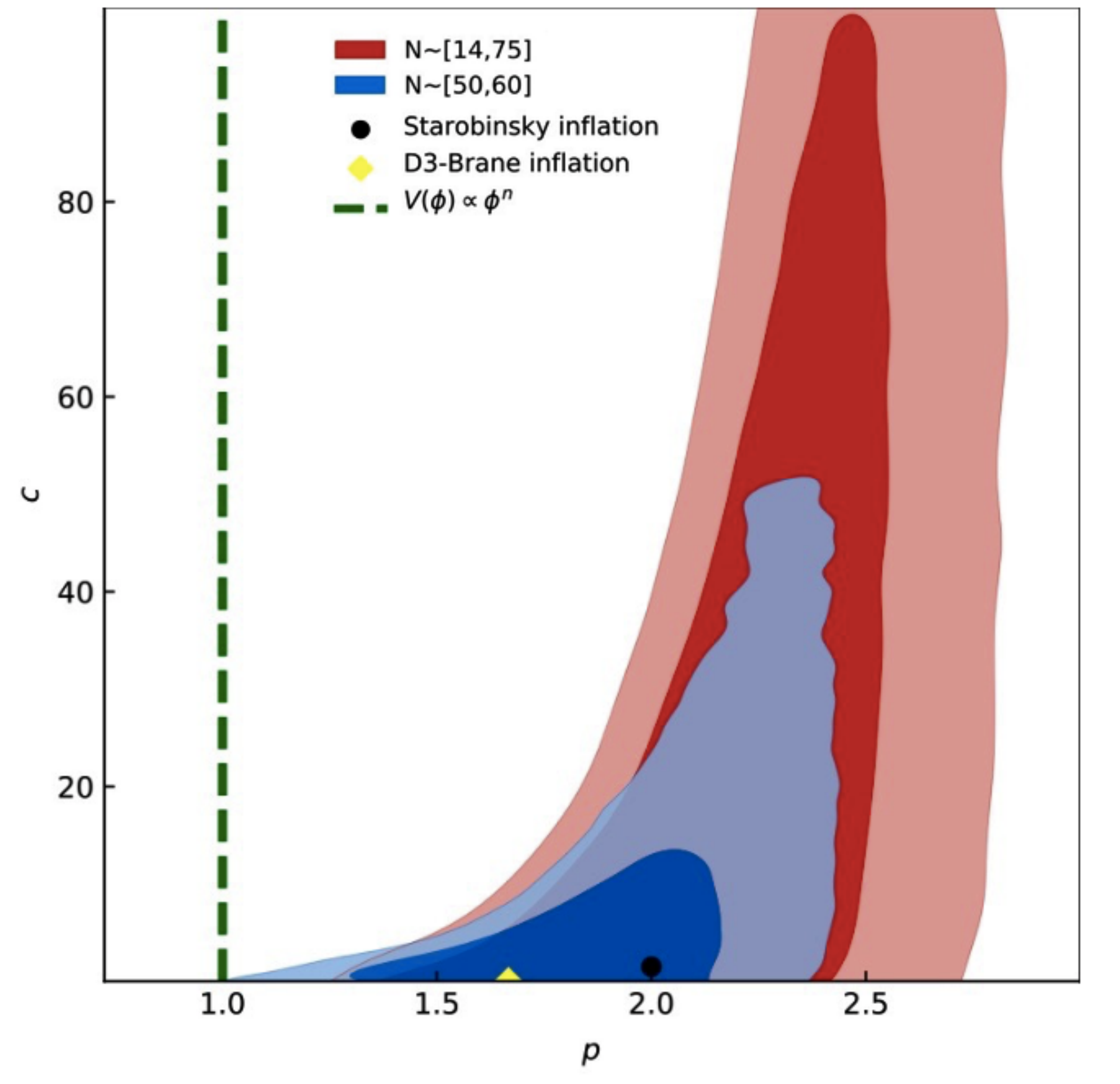}
\caption{The marginalized contour plots for parameter $p$ and $c$ at $68\%$ CL and $95\%$ CL from  Planck15+BK15+BAO+JLA datasets. The blue and red regions correspond to $N\in[50,60]$ and $N\in[14,75]$ condition, respectively. }
\label{fig:5}
\end{figure}

In this subsection, the tensor-to-scalar ratio and the scalar spectral index can be replaced by parameters $N$, $p$ and $c$ which are all taken as free parameters. Usually the range of $N$ is taken as $N\in [50,60]$, and the constraints on $p$ and $c$ read 
\m
p &=&  1.92\pm 0.27\quad(68\% \ \mathrm{CL}),\\
c &<&  33.1\quad(95\% \ \mathrm{CL}). 
\n
Here we also consider a more conservative estimation, namely $N\in [14,75]$ in \cite{Alabidi:2006qa}, and hence the constraints are slightly relaxed to be 
\m
p &=&  2.29^{+0.32}_{-0.25}\quad(68\% \ \mathrm{CL}),\\
c &<&  50.2\quad(68\% \ \mathrm{CL}). 
\n
In both cases, the models with $p=1$ corresponding to $V(\phi)\sim \phi^n$ are disfavored at more than $95\%$ CL, but the Starobinsky inflation model and brane inflation model can fit the data quite well. See the results in Fig.~\ref{fig:5}. 

\section{Summary and Discussion}

In this paper we revisit the constraints on the inflation models by adopting the latest datasets, in particular the BICEP2/Keck CMB polarization data up to and including the 2015 observing season which yields the most stringent constraint on the tensor-to-scalar ratio up to now. On the other hand, in order to break the degeneracies among some cosmological parameters, we also include the datasets of Baryon Acoustic Oscillation and the Type Ia supernovae measurements. We find that both the inflation model with monomial potential and natural inflation model are disfavored, and the inflation models with a concave potential, such as the Starobinsky inflation model, brane inflation model, hilltop inflation model \cite{Huang:2015cke,Barenboim:2016mmw,Kinney:2018kew} and so on, are preferred.  


\noindent {\bf Acknowledgments}.

This work is supported by grants from NSFC (grant No. 11690021, 11575271, 11747601), the Strategic Priority Research Program of Chinese Academy of Sciences (Grant No. XDB23000000, XDA15020701), and Top-Notch Young Talents Program of China.



\end{document}